# Automatic characterization of boulders on planetary surfaces from high-resolution satellite images

Nils C. Prieur[1,2,3] (nilscp@stanford.edu), Brian Amaro[1], Emiliano Gonzalez[1,4], Hannah Kerner[5], Sergei Medvedev[6], Lior Rubanenko[1,7], Stephanie C. Werner[2,3], Zhiyong Xiao[8], Dmitry Zastrozhnov[9,10,11], Mathieu G. A. Lapôtre[1].

[1]Department of Earth & Planetary Sciences, Stanford University, USA. [2]Center for Earth Evolution and Dynamics. University of Oslo, Norway. [3]Center for Planetary Habitability, University of Oslo, Norway. [4]Department of Geological Sciences, California State Polytechnic University Pomona, USA. [5]School of Computing & Augmented Intelligence, Arizona State University, USA. [6]Medvedev Consulting, Oslo, Norway. [7]Deptartment of Civil & Environmental Engineering, Technion, Israel. [8]Planetary Environmental and Astrobiological Research Laboratory, School of Atmospheric Sciences, Sun Yat-Sen University, China. [9]Volcanic Basin Energy Research, Oslo, Norway. [10]A.P. Karpinsky Russian Geological Research Institute, Saint Petersburg, Russia. [11]Department of Geosciences, University of Oslo, Norway.

**Abstract**

Boulders form from a variety of geological processes, which their size, shape, and orientation may help us better understand. Furthermore, they represent potential hazards to spacecraft landing that need to be characterized. However, mapping individual boulders across vast areas is extremely labor-intensive, often limiting the extent over which they are characterized and the statistical robustness of obtained boulder morphometrics. To automate boulder characterization, we use an instance segmentation neural network, Mask R-CNN, to detect and outline boulders in high-resolution satellite images. Our neural network, BoulderNet, was trained from a dataset of > 33,000 boulders in > 750 image tiles from Earth, the Moon, and Mars. BoulderNet not only correctly detects the majority of boulders in images, but it identifies the outline of boulders with high fidelity, achieving average precision and recall values of 72% and 64% relative to manually digitized boulders from the test dataset, when only detections with intersection-over-union ratios > 50% are considered valid. These values are similar to those obtained by human mappers. On Earth, equivalent boulder diameters, aspect ratios, and orientations extracted from predictions were benchmarked against ground measurements and yield values within ±15%, ±0.20, and ±20º of their ground-truth values, respectively. BoulderNet achieves better boulder detection and characterization performance relative to existing methods, providing a versatile open-source tool to characterize entire boulder fields on planetary surfaces.

**Key Points**

- We describe BoulderNet, a Mask R-CNN instance segmentation model trained to characterize boulders on the surfaces of solid planetary bodies.
- BoulderNet achieves robust performance on Earth, the Moon and Mars, outlining the boulders with high fidelity.
- BoulderNet achieves better boulder detection and morphometric characterization performances than other existing methods.

**Plain Language Summary**

Boulders are one of the most abundant features on the surfaces of solid planetary bodies. Measuring their size, shape, and orientation can tell us about how they formed as well as help select landing sites that minimize hazard to spacecraft. However, mapping boulders across large areas is a labor-intensive task that often limits the scope and robustness of boulder studies. To overcome this challenge, we trained a machine-learning algorithm to automatically outline boulders on a variety of planetary surfaces using a database of over 30,000 boulders manually mapped from aerial or satellite images of Earth, the Moon, and Mars. Our algorithm, BoulderNet, performs as well as human mappers and outperforms existing automated tools. BoulderNet is made available to the planetary community.

**1. Introduction**

Boulders are rock fragments larger than 25.6 cm (Wentworth, 1922) that can form from a wide variety of physical and chemical processes. For example, fragmentation of planetary crusts upon asteroid impacts generates extensive boulder populations (e.g., Bart & Melosh, 2010), similar to those produced by chemical and nuclear explosions or volcanic eruptions (Michikami et al., 2019). Other common sources of in situ fragmentation include thermal fracturing, frost shattering, sublimation, or any other process that generates changes in rock volume (Klemån & Borgström, 1990; Eppes et al., 2015; Attree et al., 2018). Mass-wasting events such as landslides, rockfalls, and debris flows can break down damaged bedrock into boulders (Shobe et al., 2021). Similarly, glaciers (Darvill et al., 2014), rivers and floods (Elfström, 1987), as well as storms, hurricanes, and tsunamis (Dewey et al., 2021) can erode bedrock under the action of a shearing fluid.

As the shape, size, and orientation distributions of boulders often relate to their formation process, they offer a unique opportunity to decipher the geologic history of planetary surfaces. For example, boulders have been used to identify the geological processes shaping planetary landscapes (Xiao et al., 2013; Bickel et al., 2020; Dewey et al., 2021; Pajola et al., 2021), infer the past existence of multiple glaciations on Mars (Levy et al., 2021), determine the direction of glacier flows (Darvill et al., 2014), and estimate erosion rate (Watkins et al., 2019; Basilevsky et al., 2013). Last but not least, quantitative characterizations of boulders are critical to landing-site selection and hazard assessment for landed missions (Golombek et al., 2012; Pajola et al., 2017; Golombek et al., 2017).

Individual boulder fields may include hundreds of thousands of boulders or more (Bart & Melosh, 2010; Krishna & Kumar, 2016; Pajola et al., 2019; Watkins et al., 2019), such that manual mapping of individual boulders across entire fields is highly impractical, significantly limiting the area over which boulders are typically mapped, and thus, limiting the statistical robustness of boulder morphometrics and ensuing interpretations. Automating boulder detection, mapping, and characterization is therefore key to obtaining statistically significant, high-quality boulder morphometrics over large areas and on a variety of planetary surfaces. Several boulder detection algorithms have been developed to date (Golombek et al., 2008; Wu et al., 2018; Bickel et al., 2019; Zhu et al., 2021; Hood et al., 2022; Shimizu et al., 2022) and can generally be classified into two families.

A first family of models relies on rule-based approaches, often detecting the shadows cast by boulders rather than the boulders themselves (Golombek et al., 2008; Wu et al., 2018; Hood et al., 2022). In some of these algorithms, the outlines of shadows are extracted using different segmentation techniques before being fitted to an ellipse, from which the height and diameter of an equivalent cylindrical (Golombek et al., 2012) or spherical (Hood et al., 2022) boulder are reconstructed using various geometric assumptions. Similarly, the detection routine of Wu et al. (2018) relies on the transition in pixel brightness along the illumination direction, where changes from bright to dark imply the transition from a boulder to its shadow. Whereas they do not directly use boulder shadows, Nagle-McNaughton et al. (2020) adopted a rule-based approach that exploits the contrast in pixel brightness between a boulder and its background surface to detect its outline through a variety of filtering and masking steps. Such algorithms have been tested over several locations and landing sites on Mars (Golombek et al., 2008, 2012) and on the Moon (Wu et al., 2018), and perform best on flat, horizontal surfaces as roughness elements other than boulders may also cast shadows, and topographic slopes affect shadow length. These constraints are not typically problematic when considering potential landing sites, as the latter are often picked to minimize hazards to the spacecraft and, thus, satisfy strict roughness and slope criteria (Golombek et al., 2008, 2012, 2017; Wu et al., 2018; Hood et al., 2022). However, the applicability of those algorithms may be limited in other contexts. Shadow detection to date has largely relied on the assumption of spheroidal boulders and, therefore, could not provide information about actual boulder shapes and orientations. Finally, these algorithms do not perform as well when boulders cast minimal shadows, e.g., under high illumination angles or when boulders are partially buried.

A second family of algorithms relies on deep learning approaches (Bickel et al., 2019; Zhu et al., 2021; Shimizu et al., 2022) and typically uses Convolutional Neural Networks (CNN). CNNs are powerful deep-learning techniques that learn from images of objects of interest through training and automatically classify them in new images (LeCun et al., 1989). CNNs have progressively been applied to the analysis of various planetary landforms from satellite imagery (DeLatte et al., 2019; Kerner et al., 2019; Hoeser et al., 2020a,b; Rubanenko et al., 2021). An advantage of deep-learning approaches is that boulder detection can be made size-, brightness-, and density-invariant through augmentation of the training dataset (e.g., by including images with different illumination conditions, sizes, and spatial density of boulders). For example, boulder falls have been detected globally on the Moon by identifying boulder tracks in images (Bickel et al., 2019). Whereas boulder tracks make recognition of boulders easier, boulders without tracks have also been automatically detected on the Moon (Zhu et al., 2021) and Bennu (Shimizu et al., 2022). However, the CNN architecture used by Bickel et al. (2019) and Zhu et al. (2021) only detects and outlines the bounds of boulders using rectangular bounding boxes, limiting their applicability to quantify boulder morphometrics.

Here, we develop, train, and test a CNN model to automatically detect boulders from satellite and drone images on any solid planetary body (Section 2.1). We use an instance segmentation neural network (He et al., 2017) that outlines individual boulders in a given image, enabling the extraction of boulder morphometrics. To be applicable to a wide range of high-resolution image products returned from spacecraft missions, the model uses a single greyscale band in the visible wavelengths as input. Boulder sizes, shapes, and spatial density vary considerably across planetary surfaces due to differences in terrain properties (Pajola et al., 2017), mode of emplacement (Morris & Shoemaker, 1970; Cintala & Mcbride, 1995; Krishna & Kumar, 2016), and subsequent erosional processes (Watkins et al., 2019). To develop a versatile boulder detection algorithm, the model is trained on boulder fields sampling different planetary surfaces, formation mechanisms, illumination conditions, and ranges in sizes, shapes, degradation states, and abundances. The algorithm, BoulderNet, is made freely available to the community (Prieur et al., 2023a).

## 2. BoulderNet: Methods

### 2.1 Convolutional neural network, architecture, and platform

CNNs are specialized artificial neural networks that process grid-like data, such as images, speech signals, and videos and excel at supervised learning tasks (LeCun et al., 2015). Unlike traditional neural networks, CNNs use convolutional layers to learn local patterns or features by sliding small filters across input grids. Pooling layers subsample convolutional outputs to reduce dimensionality, improve pattern recognition, and minimize computation. Each convolutional- and pooling-layer output typically passes through a nonlinear activation function (such as, e.g., ReLU, Fukushima, 1975) to increase expressiveness. During training, the loss function, which measures the deviation between modeled and expected outputs, is minimized. CNNs can be complex and require hundreds of billions of parameters to learn specific features in images.

Here, we opt to use a CNN instance segmentation architecture, similar to that used by Shimizu et al. (2022). Unlike object detection, which identifies bounding boxes around objects, and semantic segmentation, which labels each pixel in an image with an object class, instance segmentation assigns a unique ID to each within a given class, such that objects within the same class can be distinguished. Thus, this approach is best suited to detect instances of boulders and create boulder outlines of arbitrary shapes. Among potential candidate architectures, the instance segmentation Mask R-CNN neural network has achieved state-of-the-art performance on various benchmark datasets, including COCO and Cityscapes (He et al., 2017). Mask R-CNN is widely used in computer vision applications such as autonomous driving, robotics, and medical imaging (Johnson, 2018; Malbog, 2019; Jia et al., 2020; Shu et al., 2020; Fang et al. 2023). Moreover, it has previously been used for studying surface processes from satellite imagery, yielding sufficient accuracy in characterizing the morphometrics of various landforms (e.g., up to ~80% mean average precision for planetary dunes and boulders; Rubanenko et al., 2021; Shimizu et al., 2022).

Mask R-CNN performs instance segmentation to detect objects and their corresponding masks in an image (He, 2017). In a first stage, a backbone CNN extracts features from the input image using established architectures like ResNet (He et al., 2015). The number of layers in the backbone is manually tuned to avoid over- or underfitting. In a second stage, a region proposal network (RPN) scans the feature map to propose positive and negative anchors, which are parts of the image that likely contain objects or background, respectively. The RPN uses the intersection-over-union ratio (IoU, Section 2.3) to determine the boundaries of positive anchors, which are then filtered through non-maximum suppression (Neubeck & Van Gool, 2006) to create regions of interest (ROIs). A feature pyramid network (FPN) then examines ROIs in a final, third stage to determine their classification and binary pixel mask (Lin et al., 2017).

To train Mask R-CNN and make predictions, we use Detectron2, a PyTorch-based machine learning platform that provides state-of-the-art detection and segmentation algorithms (Wu et al., 2019). Detectron2 provides tools and functions to support users before, during, and after algorithm training (e.g., with image pre-processing, augmentation techniques, hyperparameter tuning, and predictions). All model parameters are stored in a setup file, which can be easily shared across platforms. Those setup files can be used towards further tuning and re-training of the model, as, for example, more manually digitized boulder outlines become available. Similarly, model weights can be easily shared. Together, model parameters and weights constitute the two inputs required to automatically detect boulders in new images.

## 2.2 Input data

### 2.2.1 Terrestrial images and ground truth

Three terrestrial boulder fields were included in our dataset, two of which were used to groundtruth model predictions. The first boulder field is located near Courtright Reservoir in the Sierra National Forest in California (Figure 1a, 37.0968°N 118.9586°W). There, boulders range from tenths of centimeters to meters in diameter (Figure 1a,b) and formed as glacial erratics (Wahrhaftig et al., 1984) during the Pleistocene glaciation of the Sierra Nevada (Moore & Moring, 2013). Boulders rest on a gently dipping (~5–10°) granitic bedrock slope (about 530×100 m) that is largely devoid of vegetation (Figure 1a). The boulder spatial density is relatively low, and the contrast between the boulders and background bedrock is relatively poor (Figure 1b). The distribution of boulders at Courtright Reservoir resembles the low-density boulder populations found at the base of degraded-crater walls on the lunar surface (Figure 1c).

Our second boulder field is located on the South Deadman volcanic dome within the Inyo-Mono crater chain in California (37.7160°N 119.0195°W; Figure 1d). Mono-Inyo craters are a 29-km long chain of craters found along the eastern Sierra Nevada between Mono Lake and Long Valley Caldera (Bailey et al., 1976; Bailey, 2004) and which formed from phreatic explosions and subsequent flows of viscous rhyolitic lava (Wohletz & Heiken, 1992; Sharp & Glazner, 1997). South Deadman volcanic dome is one of the youngest rhyolitic domes within the Inyo crater chain and is thought to have formed 650–550 years ago (Bailey, 2004; Millar et al. 2006). A ~500×1,000 m portion of the larger South Deadman dome (~1,350×1,350 m) was investigated in the field along its northern flank. There, the diameters of rock fragments range from centimeters to decameters and are, on average, larger than at Courtright reservoir. No vegetation was observed on the rugged, rubbly dome. Boulder density is very high, with many boulders overlapping (i.e., they are imbricated; Figure 1e). This field location is a good analog for, e.g., boulder-rich asteroids such as Bennu (Figure 1f), Itokawa, and Ryugu (Noguchi et al., 2010; Michikami et al., 2019; Walsh et al., 2019; Michikami & Hagermann, 2021; Zhu et al., 2021; Arakawa, 2022).

The third location, Stonegarden (area of ~1 km$^2$), is located in the Altaï mountains along the Katun River in Russia (50.4764°N 82.6263°E, Figure 1g). One proposed origin of this boulder field is that of boulder deposition by Quaternary megafloods, resulting from the catastrophic drainage of ice-dammed lakes upstream in the Chuja, Kuraj, and Uimon depressions (Herget, 2005; Reuther et al., 2006). However, the catastrophic nature of boulder deposition has been debated, and others favor a glacial origin (e.g., Okishev, 2011; Zykin et al., 2016). The size of deposited rock fragments ranges from a few centimeters to tenths of meters (Figure 1h). The spatial density of boulders is intermediate between those at the Courtright Reservoir and South Deadman dome. In contrast to the other locations, grass and bushes are present in the study area. Whereas vegetation leads to a higher contrast between boulders and their substrate in images, bushes and boulders have similar shapes, providing a challenge for boulder detection in grayscale images. Therefore, we expect the precision of the algorithm to be lowered by a few percent. The spatial density of boulders at Stonegarden is similar to, e.g., that observed around Chryse Planitia on Mars (Figure 1i).

Aerial images were acquired at all three locations using a DJI Phantom 4 drone and processed into orthomosaics using Agisoft Metashape software. Resulting orthomosaics have resolutions of either 3 or 6 cm/pixel (Table 1). Boulders with sizes larger than 5–10 pixels were found to be fully resolved such that boulders with diameters > 15–30 cm or > 30–60 cm are expected to be detectable depending on pixel resolution. In addition, ground-truth measurements were conducted at the Courtright Reservoir and South Deadman dome through manual size and shape measurements of 20 and 12 randomly selected boulders, respectively, spanning decimeter-to-meter-scale sizes. For each boulder, GPS location, short and long axes, boulder orientation, and dip along the long axis were recorded. Furthermore, evaluated boulders were marked with orange tape to be easily identifiable in drone mosaics. Ground-truth data allows us to (1) quantify the uncertainty associated with manual mapping of boulders from aerial or satellite imagery (section 3.1), and (2) evaluate the performance of the model both in identifying boulders and in quantifying their sizes and shapes (section 3.2).

**2.2.2 Lunar and martian images**

In addition to terrestrial boulders, our dataset also includes boulders outlined from images of lunar and martian surfaces. On the Moon, we used images acquired by the Narrow Angle Cameras (NACs) onboard the Lunar Reconnaissance Orbiter (LRO) (Robinson et al., 2010). The NACs are monochrome narrow-angle line-scan imagers, which return images with ~25–100 cm/pixel resolution depending on mission phase (Robinson et al., 2010). NAC images were processed to level-2 products, following guidelines provided by the U.S Geology Survey (USGS) and the digital image processing software package ISIS3 (Jason et al., 2023), including ground positions, radiometric calibration, echo correction, and projection to a local equirectangular coordinate system. Original illumination conditions were preserved (no stretching). Repeat LRO passes allow us to investigate the impact of illumination conditions on model detection performances (section 4.2). On Mars, we used images from the High-Resolution Imaging Science Experiment (HiRISE) camera onboard the Mars Reconnaissance Orbiter (MRO) (McEwen et al., 2007). HiRISE images have a pixel resolution between 25 and 130 cm/pixel, and were pre-processed to level 2 products (i.e., ingestion of raw camera geometry, radiometric calibration, channel stitching, noise removal, map projection, and correction of tonal mismatches) using the MarsSI online platform (Quantin-Nataf et al., 2018).

**2.2.3 Data conditioning**

**2.2.3.1 Labeling**

To homogenize the quality and consistency of labels while also maximizing the size of the dataset, we followed the pipeline of Zamir et al. (2019). First, we adopted a precise definition of what constitutes a mappable boulder, developed annotation guidelines, and trained three human boulder annotators. Then, annotators were assigned labeling tasks on a small number of images, and produced labels that were quality checked as a group. Annotation guidelines were then refined iteratively through a series of labeling and quality-check exercises until guidelines were comprehensive enough to account for most challenging cases and mapping by all annotators was deemed consistent. All annotators used the Geographical Information System (GIS) QGIS software (QGIS Development Team, 2023) to outline individual boulders as polygon shapefiles.

The adoption of a clear definition of a boulder, which can be upheld across multiple annotators, and that is tailored to the model architecture (Mask R-CNN) and technique (instance segmentation) used, is key to building a high-quality dataset. Whereas many well resolved boulders are relatively easy to identify in an image, consistent mapping of boulder outlines may become challenging for a variety of reasons, including their geomorphic setting, size, spatial density, contrast against their substrate, and illumination conditions. For example, boulders generated by asteroid impacts can be buried to variable degrees by ejecta just after emplacement. Furthermore, in situ fragmentation generates boulders that often have a similar tone to that of their substrate. As all challenging cases encountered in images cannot be anticipated prior to mapping, an iterative procedure consisting of mapping exercises, quality checks, and group discussions was adopted. The following definition of a mappable boulder resulted from this iterative process: "A boulder consists of any resolvable rock fragment with a surface area larger than 5×5 pixels and that can be confidently differentiated from its substrate. A boulder's outline can be traced with a single simple polygon." Further details about our annotation and review guidelines are provided in the Supplementary Information.

**2.2.3.2 Data split**

A total of 39,540 boulders, originating from a total of 953 512×512-pixel tiles from 23 high-resolution images are included in our database (section 2.2.3.3). All images used, the number of tiles mapped, the number of boulders mapped, pixel resolution, and incidence angles are summarized in Table 1. Distribution statistics of image tiles, resolution, incidence angles, and percentage of area covered by boulders for each planetary body are summarized in Figure 2a–d. About 69% of all tiles in the database are from the lunar surface. All mapped lunar boulders are of impact origin and were mostly mapped within 1–2 crater radii away from the crater center, where the density of boulders is the highest. In contrast, terrestrial boulders represent 26% of all image tiles and are of glacial, phreatomagmatic, or flood origins. Finally, image tiles from the martian surface represent 5% of our database and comprise impact-generated boulders and, most likely, boulders formed from other unconstrained mechanisms.

We consistently sought to use the highest resolution images available to map boulder populations to the smallest possible size (most images have resolutions < 80 cm/pixel; Figure 2b). The majority of selected images have incidence angles of 25–60° to avoid blooming or shadow obstruction effects (Figure 2c; Supplementary Information Text S1). Boulder areal density is < 10% in most tiles (Figure 2d), with the exception of tiles covering the rim of some of the freshest impact craters (e.g., M139694087LE and M143954642, with areal density up to 13%), martian boulders near Chryse Planitia (e.g., up to 25% for ESP_017355_2260 and ESP_028537_2270) and boulders of volcanic origin (e.g., South Deadman dome, > 70%). The vast majority of mapped boulders have aspect ratios (defined as the ratio of their long to short axes) < 3 (Figure 2e; section 2.4).

To evaluate model performance, tiles were split into training (80%), validation (15%), and test (5%) datasets. The training dataset is used for training Mask R-CNN, the validation dataset to tune model hyperparameters, and test boulders are used as an independent dataset to evaluate how the model generalizes to yet unseen images. Figure 3 illustrates the data split for 3 of the 23 images in the database.

**2.2.3.3 Pre-processing, training, and post-processing pipelines**

Original satellite images were tiled to satisfy memory constraints during training. Digitized boulder outlines were next converted into a format that can be imported into the Detectron2 detection and segmentation platform, and each boulder was assigned a unique ID, linked to the ID of the tile it originated from. To speed up the training process, we applied transfer learning (Bozinovski and Fulgosi, 1976), and used a version of Mask R-CNN with a ResNet-50 backbone that was pretrained on the COCO dataset (Wu et al., 2019). In addition, we used a large number of augmentations (e.g., Lalitha & Latha, 2022), including techniques such as rotations, random zooming, scaling, brightness, contrast adjustments, elastic transforms, and more to artificially increase the number of image tiles and data diversity, and avoid potential overfitting. Training was conducted using an AMD Ryzen 9 5900X 12-core processor and a NVIDIA RTX A4000 graphic card, with a cycling 0.0002–0.002 learning rate (Smith, 2015), a batch size of 4, for a total of 320 epochs. Both training and validation datasets were used to decide when sufficient training was achieved.

Upon training, several post-processing steps were applied to further improve model performance. Notably, we used the Test Time Augmentation (TTA) technique (e.g., Matsunaga et al. 2017, Ayhan et al., 2018, Wang et al., 2019). We applied 8 rotations and reflections (Dummit and Foote, 2004), resulting in 8 augmented images per original image. Boulder outlines were then predicted in each of the augmented images, and overlapping boulders were removed using a non-maximum suppression (NMS) algorithm (Neubeck & Van Gool, 2006; Girshick et al., 2014, Girshick, 2015). The NMS algorithm retains the most confident prediction from all overlapping boulder outlines and filters out the others, providing an efficient approach to combine predictions made from all 8 augmented images. Further significant improvement was achieved by artificially increasing the size of tiles from 512×512 to 1024×1024 pixels using bilinear interpolation, improving prediction performance for the smallest boulders. Furthermore, image tiling leads to challenges in detecting boulders near tile edges. To circumvent this issue and enhance detection rate, we repeated the tiling process a total of six times, staggering tile boundaries each time: (i) no shift applied: (0,0) along x- and y- image coordinates; (ii) a (256, 256) shift is applied; (iii) a (256, 512) shift is applied; (iv) a (512, 256) shift is applied; (v) a (0, 256) shift is applied; and (vi) a (256, 0) shift is applied. Once boulders are detected across all six tiling configurations, only boulders with centers within the inner 62.5% of the tile are kept to favor selection of the most confident detections and allow for a small amount of overlap between boulder predictions. The non-maximum suppression technique is then again used to remove overlapping predictions. The prediction parameters available within the Mask R-CNN model setup (Wu et al., 2019) can also be fine-tuned to improve model performances. In particular, the *SCORE_THRESH_TEST* threshold parameter (Supplementary Information Table S1) controls the confidence level that dictates which predictions are deemed valid. Whereas it could be held constant across all planetary bodies (Supplementary Information Table S1), we found that model precision and recall are improved by ~5–10% when tuned for a specific image. Final boulder outlines are then stored as shapefiles, which can be easily imported into QGIS for inspection of results and data analysis.

**2.3 Evaluation of model performance**

To evaluate the detection performance of the model (after post-processing steps), we calculate the model precision and recall as:

$$\text{precision} = \frac{\text{true positives}}{\text{true positives} + \text{false positives}}, \quad (1)$$

and

$$\text{recall} = \frac{\text{true positives}}{\text{true positives} + \text{false negatives}}. \quad (2)$$

In order to decide whether a boulder prediction is considered as true or false positive we use the Intersection-over-Union (IoU), which is a metric that quantifies the overlap between a prediction and the actual outline of a boulder (Jaccard, 1912),

$$\text{IoU} = \frac{A_{\text{inter}}}{A_{\text{union}}}, \quad (3)$$

where $A_{\text{inter}}$ is the surface area of the intersection between the actual and predicted boulder outlines (i.e., the overlap), and $A_{\text{union}}$ is the surface area of the union of the actual and predicted boulder outlines. Therefore, the larger the IoU, the greater the overlap, and thus, the better the prediction is. To differentiate valid from non-valid predictions, a specific overlap threshold is often specified. For example, the $\text{IoU}_{50}$ reports performance when only predictions with $\text{IoU} \geq 0.5$ are considered true positives. Increasing the IoU threshold results in stricter performance criteria. False negatives can then be calculated by subtracting true and false positives from the total number of actual boulders. All performance metrics (Equations 1–2) are computed for both validation and test datasets. Throughout the manuscript, "ground truth" is used to refer to in situ measurements by geologists on the ground, "manual" for manually outlined boulders in images, and "actual" refers to whichever of ground truth or manual measurements are used as the control dataset to which automatic detections are compared.

## 2.4. Model hyperparameters

Optimizing model performance requires tuning many model hyperparameters. To that end, we modified the model setup one parameter at a time, trained the algorithm, and evaluated model performances on the validation dataset. A description of the most important parameters of the best model is given in Supplementary Table 1 (Supplementary Information). Three main changes were found to significantly improve model performance: (i) the level of use of augmentations, (ii) anchor sizes, and (iii) the use of a cyclical learning rate (Smith, 2015). Supplementary Table S2 summarizes the average precision and recall for $IoU_{50}$ for different model setups tested. The best precision and recall were achieved when Mask R-CNN prediction parameter *SCORE_THRESH_TEST* was tuned for each location (Section 2.2.3.3), ensemble of models (both 512×512 and 1024×1024 tile sizes) were used, and augmentations (including TTA) were applied. This version of the model was adopted for the remainder of this study.

## 2.5 Boulder morphometrics

Predicted boulder outlines are next used to characterize boulder morphometrics. First, we densify the number of vertices along boulder outlines to achieve a regular vertex spacing of half a pixel. An equivalent boulder diameter is then calculated as the diameter of a circular boulder of area ($A$) equal to that of the true boulder outline, $D = 2\sqrt{A/\pi}$. Next, an ellipse is fitted through the densified vertices, from which the boulder's long ($a$) and short ($b$) planform axes are measured. Finally, boulder orientation is calculated from the fitted-ellipse's long-axis azimuth for boulders with *a/b* > 1.35.

## 3. Results

## 3.1 BoulderNet detection performance

First, we quantify the average precision (AP) and recall (AR) of the model on the test dataset for IoU ≥ 0.50 (similar trends are observed for IoU ≥ 0.75; Tables 2–3). BoulderNet predictions yield consistently higher precision and recall values for terrestrial and lunar boulders (which constitute the majority of the training data; Figure 2e), e.g., with $AP50_{Earth}$ = 74–87% and $AR50_{Earth}$ = 61–85%, and $AP50_{Moon}$ = 66–70% and $AR50_{Moon}$ = 53–73%. Predictions for martian boulders are reasonable, with $AP50_{Mars}$ = 38–71% and $AR50_{Mars}$ = 18–68%, despite only representing 5% of the training dataset. The worst performance is observed for HiRISE image ESP_025650 (Table 3), which reveals a surface dominated by many small partially buried boulders, which are close to the resolution limit of the image for detecting boulders with confidence. Across all studied planetary bodies and 23 locations in our database, average values of AP50 = 72% and AR50 = 64% are achieved. To assess any performance dependency on boulder size, boulders are split into three populations (Table 3) – small ($6 \times 6 \leq A < 16 \times 16$ pixels), medium ($16 \times 16 \leq A < 32 \times 32$ pixels), and large ($A \geq 32 \times 32$ pixels). Performance metrics do not vary systematically with boulder size across sites.

Second, we compare the cumulative frequency distributions of boulder sizes (using boulder equivalent diameter), or BSFDs (Boulder Size-Frequency Distributions), derived from model predictions and manually digitized boulder outlines (Figures 4a, 5a, and 6a). Examples of model predictions from three randomly picked test image tiles are shown for the Earth, Moon, and Mars in Figures 4b–d, 5b–d, and 6b–d, respectively. With an overall AR50 of 64%, model BSFDs tightly follow those obtained from manual measurements. The largest discrepancies between manual and model outlines are observed for the largest boulders at South Deadman dome (Figures 4a,c) due to its very high areal density of boulders, and for martian boulders (Figure 6a), which are challenging even for human mappers (section 4.1). Importantly, the slopes of manual and model BSFDs are roughly parallel in all locations and across boulder sizes, including where the BSFDs roll over (either due to mapping completeness or preferential removal/weathering of smaller boulders), confirming the overall absence of any important and systematic size-bias in model performance relative to expert human mappers.

## 3.2 Comparison with ground truth and manual mapping

To test the reliability of boulder morphometrics extracted with BoulderNet, we first compare our ground-based measurements of boulder size, shape, and orientation with those obtained from manually mapped boulder outlines in drone images (Figure 7a–c). Ground-based measurements were projected onto a horizontal plane with the help of a constant $10^{o}$ slope to correct for both individual dips in boulders and the local topographic slope (5–15 $^{o}$ slopes at both sites). The largest discrepancies are found for the most elongated boulders (aspect ratio $\geq$ 1.6) at Courtright Reservoir, likely due to the larger uncertainties in the ground measurement of boulder geometries along elongated dipping boulders (Figure 1a). Overall, boulder diameters, aspect ratio, and orientation derived from digitized and predicted boulders cluster tightly along 1:1 lines and remain within $\pm15\%$, $\pm0.20$, and $\pm20^{o}$ of their ground-truth values, respectively.

Second, we compare the same suite of boulder morphometrics as obtained from manual and automated mapping (Figure 7d–f). To compare each manually mapped boulder to its exact counterpart in the automated dataset, only predictions with IoU $\geq$ 0.75 are considered (for a total of 2,410 boulders). Again, boulder diameters, aspect ratios, and orientations largely remain within $\pm15\%$, $\pm0.20$, and $\pm20^{o}$ of each other, respectively, and aspect ratio appears to be the most challenging property to quantify, likely due to the abundance of small boulders (and the relatively large error imparted by single-pixel imprecision on those). In general, more scatter is to be expected in the morphometrics of small boulders.

Finally, we compare the overall morphometric distributions of manually mapped and predicted boulder outlines at Courtright Reservoir (Figure 7g–i). About 70% of manually mapped boulders are detected for the lowest diameter bins (0.12–0.18 and 0.18–0.24 m), and detection rate increases as boulder diameter increases (Figure 7g). Whereas the absolute number of boulders is not recovered in full, overall distributions in boulder diameters, aspect ratios, and orientations are well reproduced.

## 4. Discussion

### 4.1 Baseline comparison: Variance among human mappers

To further evaluate the performance of BoulderNet, we sought to compare it with the inherent variance among multiple human mappers. We selected three locations of variable mapping difficulties (different environments, image resolutions, boulder densities, and formation mechanisms) located on the Earth, Moon, and Mars. Three different human expert annotators were given the task to digitize the outlines of all boulders within their assigned tiles, without any additional guidelines. Only boulders with areas greater than 36 pixels are kept for this comparison (Figure 8; Table 4). The main sources of disagreement among mappers include inconsistent delineation of shadow boundaries, non-systematic mapping of boulders near the areal threshold limit, and disagreements in the outlines of partially buried or low-contrast boulders. We found that, intuitively, agreement decreases as mapping complexity increases, with precision and recall (AP50 and AR50) values ranging from 53–99% at the Courtright Reservoir to 30–92% on the lunar and martian surfaces. We emphasize that precision and recall values obtained with BoulderNet (Table

2) are always either better or within a few percent of the average agreement between mappers (Table 4). Thus, BoulderNet performs, in average, equally well to an expert human mapper. Moreover, it ensures consistency across environmental and illumination conditions. Last but not least, BoulderNet overwhelmingly outperforms human mappers in terms of efficiency, only requiring 30–60 minutes (for computers equipped with a Graphics Processing Unit, or GPU) to map boulders over whole NAC (~13,000x56,000 pixels) or HiRISE (~28,000x51,000 pixels) images.

**4.2 Influence of pixel resolution and illumination conditions**

To investigate the influence of both pixel resolution and illumination conditions on BoulderNet's detection rate, we conducted pair comparisons of three images acquired over the same location along the northern rim of the Censorinus impact crater. The first two images (Figure 9a–b) have similar incidence angles (67–70°) but pixel resolution is 23% lower in the second image. In contrast, the second and third images (Figure 9b–c) have similar pixel resolution but incidence angle varies from 67 to 39°. Comparing the first two images (Figure 9a–b), we confirm that higher pixel resolution leads to the detections of a larger number of small-size boulders. Comparing the second and third images (Figure 9b–c), we find a good detection agreement for the largest boulders, and more small boulders being detected under higher incidence angles. Under higher incidence angles, boulders tend to cast longer shadows – a feature which the model may learn to recognize and thus, may help in the detection of the smallest boulders.

A cluster of boulders in the study area yields discrepant predictions across the three images (insets in Figure 9). Notably, a large boulder is not detected by the model in one case, even though a smaller nearby boulder is. An inspection of raw predictions made for M1123377165RE (Figure 9a) revealed that the large boulder was actually detected but subsequently removed in post-processing by NMS (Non-Maximum Suppression) because it was deemed to be a duplicate, and the smaller boulder that overlapped with it had a higher prediction confidence. Lowering the NMS threshold could help circumvent this caveat, although the currently adopted threshold was found to optimize removal of duplicate predictions. Alternatively, other approaches to remove duplicates could be used, such as Soft-NMS (Bodla et al., 2017), Non-Maximum Weighted (Zhou et al., 2017), or Weighted Boxes Fusion (Solovyev et al., 2019). Furthermore, a boulder's own roughness may cast complex shadows on the boulder's surface, such that slight differences in illumination conditions may significantly change its appearance (e.g., one large boulder may "look like" two smaller boulders; Figure 9b–c).

Boulder size-frequency distributions (BSFDs) extracted from those three images (Figure 10) further highlight the impact of pixel resolution and illumination conditions. Despite having the highest number of detections, some of the largest boulders are missing in M1218745643LE (Figure 9a), resulting in an underestimated frequency of large boulders (dashed vs. solid purple lines in Figure 10). Images with coarser resolutions (M11123377165RE and M1151645127RE, Figure 9b–c) yield very similar BSFDs despite their significantly different illumination conditions. Notably, they yield a good match to the manual BSFD for the largest boulders, and overshoot the manual BSFD for boulder diameters smaller than 10 m. This discrepancy results from a combination of large boulders being occasionally detected as multiple small boulders and the tendency to overestimate boulder size close to the resolution limit.

**4.3 Comparison with other existing boulder-mapping approaches**

Finally, we compare the performance of BoulderNet with that of two other rule-based boulder detection algorithms – the Nagle-McNaughton (NM) GIS-based boulder detection algorithm (Nagle-McNaughton et al., 2020) and the Martian Boulder Automatic Recognition System (MBARS; Hood et al., 2022). The comparison is made against manually outlined boulders mapped on a small 160×160 m portion of HiRISE image TRA_000828_2495, the test area over which Hood et al. (2022) compared their approach to the NM algorithm (Figure 11).

The NM method (Nagle-McNaughton et al., 2020) exploits the sharp contrast in brightness between a boulder and its background to detect the boulder's edge. The prediction shapefile for the NM method was obtained using both a conservative and a more liberal brightness thresholds, which was selected by Hood et al. (2022). Upon visual inspection, predictions made using a conservative threshold were found to yield better results and is thus used here for our comparison. We further applied an area threshold of 10 pixels to remove numerous small false negatives. We find that, over the test area, many predictions made by the NM method (Figure 11b) outline shadows cast by boulders or small-scale topography rather than boulders themselves, yielding low precision and recall values of 1 and 2%, respectively. As suggested by Hood et al. (2022), we interpret this relatively poor performance to result from the drastically different boulder-background contrast in the test area relative to where the NM method was originally calibrated and tested.

MBARS, in turn, detects shadows and models its corresponding boulder as a half-buried spheroid (Hood et al., 2022). As a result, detected boulders inherently have circular planform geometries. The only model input is the maximum brightness threshold used to separate shadow from non-shadowed areas. The shapefile was provided by Hood et al. (2022) and corresponds to a model where the brightness threshold was specifically tuned for the test area. MBARS detects most manually mapped boulders, with AP50 and AR50 values of 56% and 47%, respectively. Where discrepancies exist, they often originate from the simple geometric model used to approximate boulder shapes from their shadows. Whereas approximating boulders as half spheroids seems adequate for relatively equant boulders ($a/b \sim 1.0 - 1.2$), sizes appear to be consistently overestimated for more elongated boulders, and the location of their centers to be shifted.

Third, we detected boulders in the test area using BoulderNet. We note that TRA_000828_2495 was not included in the training dataset, and the prediction parameter *SCORE_THRESH_TEST* was tuned to the test area in order to obtain the best performances. We also underline that only predictions with surface areas above 10 pixels are selected, as done for the NM and MBARS methods. In the test area, BoulderNet achieves AP50 and AR50 values of 77% and 82%, respectively – the best performance of all three approaches. We further note that those values are likely lower bounds as an inspection of false positives revealed that at least 10 of those predictions were actual boulders that had not been manually digitized (due to their true area being just below 10 pixels). One meter-scale crater and one polygonal ground pattern were misinterpreted as boulders by the model. Small craters are often similar in appearance to boulders, rendering them particularly prone to yielding false-positive detections. However, contrary to boulders, impact craters impart negative topography and can thus often be differentiated from boulders based on illumination direction. Whereas illumination direction is not currently used as an input to the model, updates to the model will be released through new model versions posted in all appropriate repositories (see Open Research statement). False positives can be relatively numerous beyond two crater radii away from the center of fresh impact structures, where fields of secondary craters about the same size as the largest boulders are found. Therefore, additional processing may be needed when applying the current model implementation around fields of meter-scale secondary craters. Finally, a few boulder sizes were overestimated, leading to false positive predictions. The bulk of these overestimates arise from the pixelized nature of predicted outlines (i.e., pixel-scale accuracy).

Next, we compare boulder morphometrics as extracted from manual mapping and the three automated techniques (Figure 12). Comparing BSFDs across all four datasets (Figure 12a), we find that the NM method yields the most detections overall, due to the large number of false positives in the test area. MBARS overestimates the frequency of boulders with diameters > 2 m but underestimated that of smaller diameters (Figure 12a–b). Hood et al. (2022) described this bias as a potential shortcoming of the technique resulting from how touching and overlapping shadows are either split or merged, though some discrepancy also likely arises from overestimated boulder sizes owing to geometric assumptions for elongated boulders (as evidenced by the long-tailed distribution of equivalent boulder diameter; Figure 12b). BoulderNet also overshoots the manually derived BSFD; however, the slope and shape of the BoulderNet BSFD roughly mimics that of manually digitized boulders across all sizes. The number of detections for boulders > 2 m ($n = 23$ vs. $n = 19$) and < 2 m ($n = 173$ vs. $n = 172$) are overall consistent. Predicted boulder diameters (Figure 12b) display a very similar distribution to that of manually digitized boulders, with a pronounced peak for the bin range of 1.50±0.25 m. We observe a smaller number of detections for the smallest bin range, which corresponds to boulders with dimensions < 4–5 pixels, close to the algorithm's detection limit. Unlike MBARS, which does not allow to constrain the shape and orientation of boulders, BoulderNet adequately replicates the distributions of boulder aspect ratios and orientations in the test area (Figure 12c–d).

## 5. Conclusion

We used Mask R-CNN, an instance segmentation model, to detect and outline boulders at several locations on Earth, the Moon, and Mars. We trained the model with over 30,000 manually digitized boulders distributed across 753 image tiles. Our model, BoulderNet, achieves high performance metrics compare to previous boulder detection algorithms, with errors comparable to the variance observed between human mappers. Boulder size, shape, and orientation extracted from BoulderNet predictions were benchmarked against field measurements, yielding values within ±15%, ±0.20, and ±20º of values measure on the ground, respectively. BoulderNet has multiple advantages relative to existing non-deep learning techniques, not only yielding better detection performances, but also allowing for the characterization of robust boulder morphometrics from boulder outlines. The current model performs well across a range of planetary surfaces, environments, illumination conditions, and boulder densities, and could easily be expanded for use beyond the Earth, Moon, and Mars through the inclusion of more training data, e.g., from asteroids. Furthermore, model weights and parameters could be used for transfer learning as more and more powerful instance segmentation architectures become available in the future. Altogether, BoulderNet provides a versatile tool to characterize boulders from satellite imagery, with strong potential to help planetary scientists decipher the geological history of planetary surfaces and assess landing hazard and terrain traversability for upcoming landed missions.

**Open Research**

The configuration file and model weights of our best-trained model - the two main required inputs to detect boulders automatically at new locations, as well as the training, validation, and test datasets are available Prieur et al. (2023b). Those files are accompanied by detailed descriptions and tutorials and the entire code used in the study (Prieur et al., 2023a). HiRISE images of Mars are available via the University of Arizona & NASA PDS (McEwen, 2007). LROC NAC images of the Moon are available from the Arizona State University & NASA PDS (Robinson, 2011). Images of the three terrestrial sites were acquired by the authors using a DJI Phantom 4 drone and are included in the Zenodo archive (Prieur et al., 2023b), along with all level-2 processed images.

**Acknowledgements**

We thank Dr. Bradley Thomson for editorial handling, and Dr. Don Hood and an anonymous reviewer for constructive comments. We thank Michael Hasson for assistance in the field, Vladimir Zykin and his research group for sharing field data, David Völgyes for insights on post-processing, and Don Hood for providing prediction shapefiles from both the NM and MBARS detection algorithms for HiRISE image TRA_000828_2495. NCP acknowledges support from the European Commission (101030364) and Research Council of Norway (328597) through the MSCA-2020 Individual Global Fellowship hosted at Stanford University and the University of Oslo. This work was also supported by the Research Council of Norway through its Centres of Excellence scheme, project numbers 223272 (CEED) and 332523 (PHAB).

**Figures**

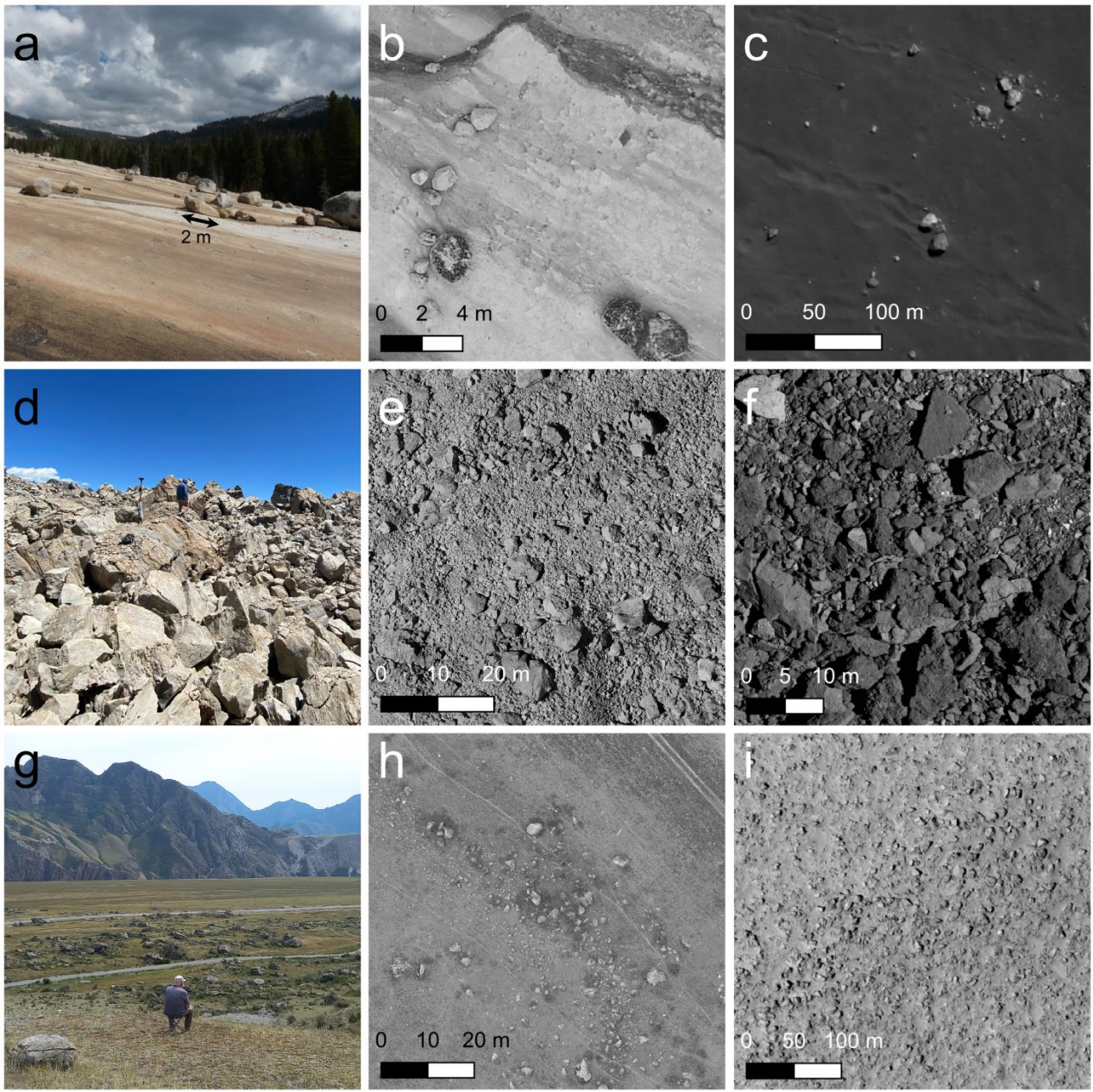

**Figure 1.** (a) Field and (b) drone pictures of glacial erratics near Courtright Reservoir (California, USA). (c) Lunar Reconnaissance Orbiter NAC (Robinson et al. 2010) image (M126214862LE) showing boulder falls within Scaliger crater on the Moon (27.1°S, 108.9°E). (d) Field and (e) drone pictures of Obsidian volcanic dome (CA, USA). (f) Osiris-REX PolyCam (Rizk et al, 2018) image of the surface of asteroid Bennu (credit: NASA/Goddard/University of Arizona). (g) Field and (h) drone pictures of a boulder field at Stonegarden in the Altaï mountains of Russia. (i) Mars Reconnaissance Orbiter HiRISE (McEwen et al., 2017) image (ESP_017355_2260) of boulders near Chryse Planitia (45.82°N, 16.47°E). Note the similarities between the terrestrial (a–h) and extra-terrestrial boulder fields (c,f,i). Humans for scale in (d,g).

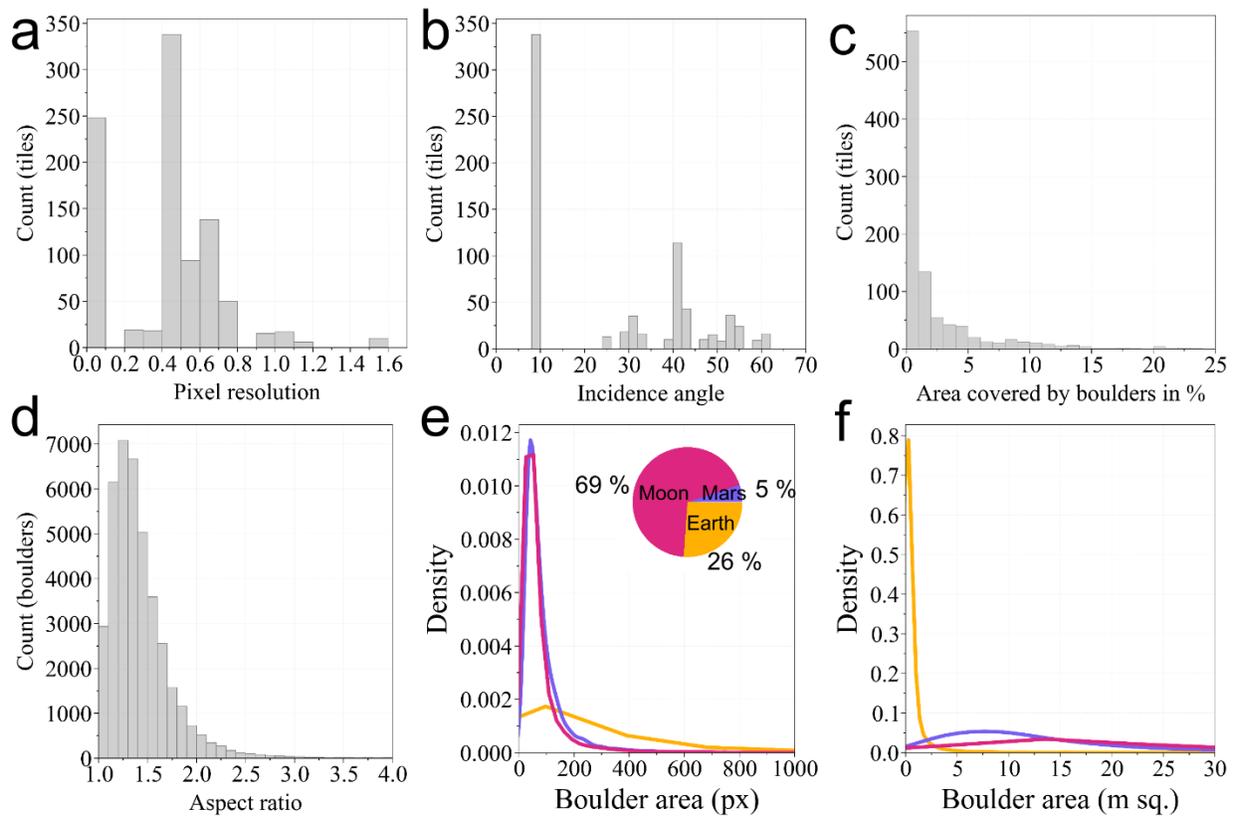

**Figure 2.** Distribution of mapped tiles (as of 25/04/2023) by (a) pixel resolution, (b) incidence angle, and (c) boulder density (expressed in % of the area covered by boulders). (d) Distribution of boulder aspect ratios in the database. Probability density functions of (e) boulder areas in pixels (inset shows distribution of tiles per planetary body and serves as a color legend for e-f), and (f) in square meters.

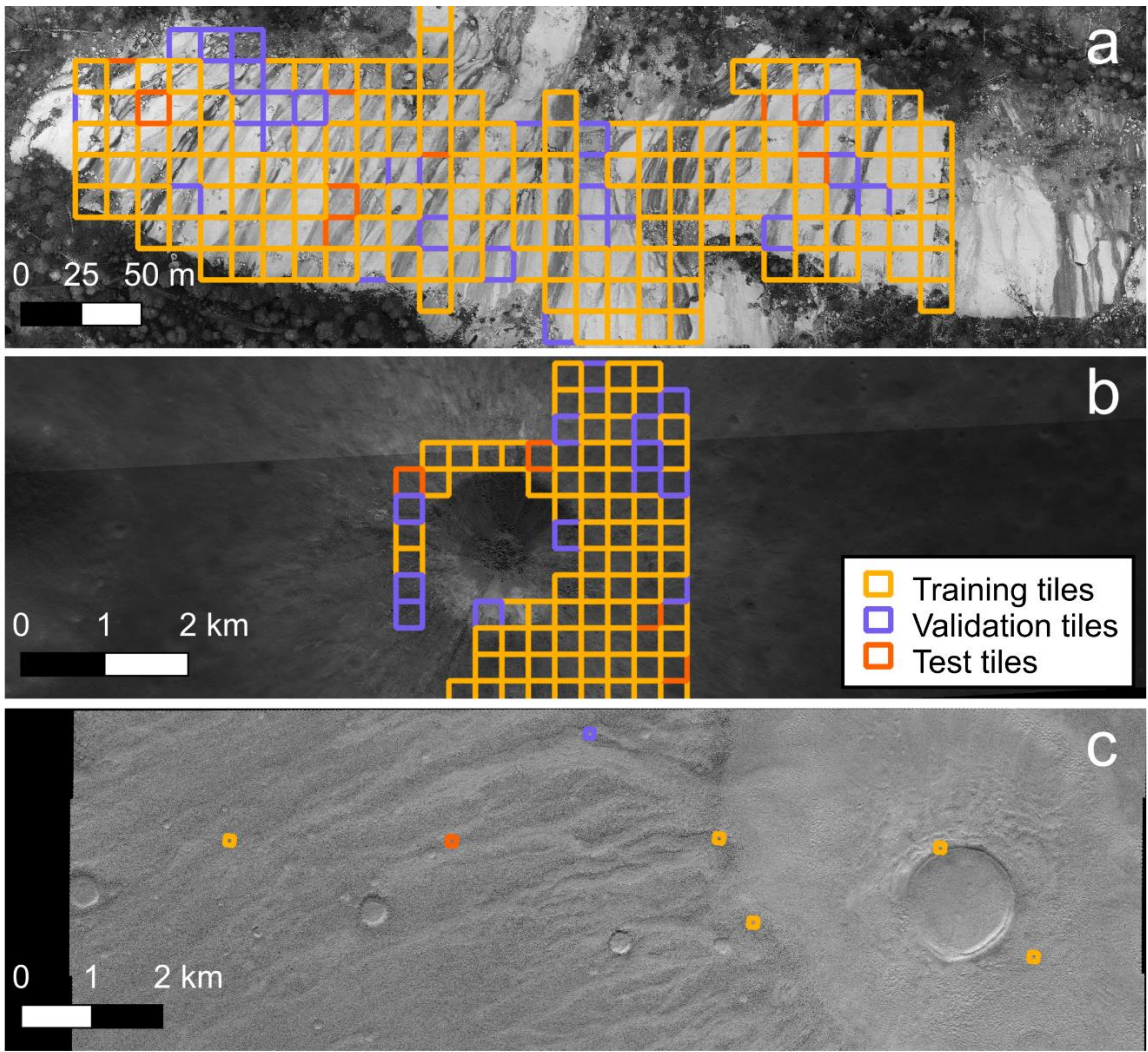

**Figure 3.** Example of a training-validation-test 80-15-5% split of mapping tiles (a) on Earth (Courtright reservoir, California, USA), (b) the Moon (M1221383405), and (c) Mars (ESP_028537_2270).

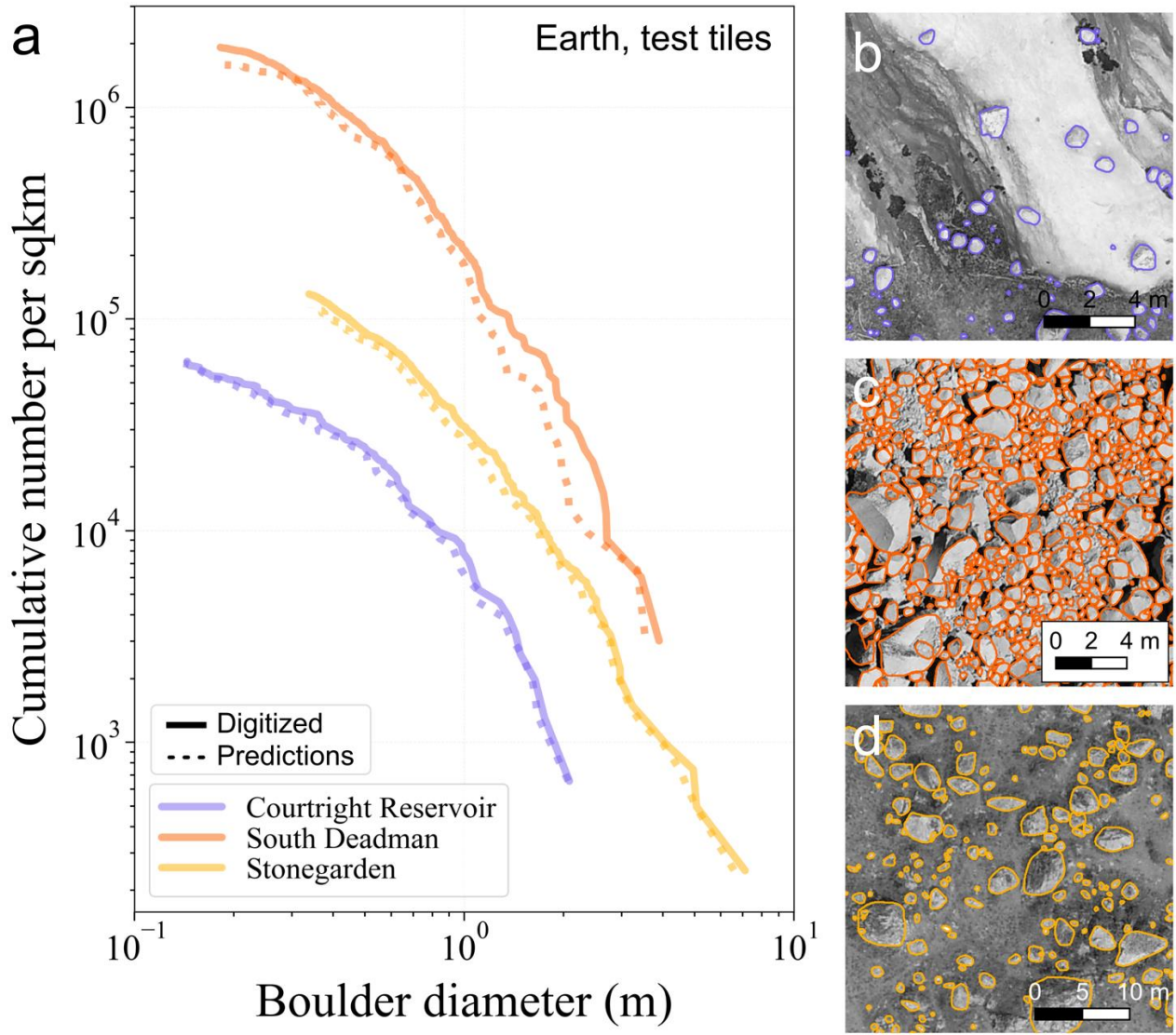

**Figure 4.** (a) Boulder-size frequency distribution (BSFD) curves for manually digitized (solid lines) and predicted (dashed lines) boulders at the terrestrial locations shown in (b-d). Only boulders with surface areas > 6×6 pixels are shown (similar to manual mapping methodology, but with one added pixel in each direction to minimize potential threshold effects); variability on the smallest mapped boulders results from variable pixel resolution among sites. Example close-up predicted boulder outlines at a randomly picked test tile at (b) Courtright Reservoir, (c) South Deadman volcanic dome, and (d) Stonegarden. Supplementary file Animation S1 shows a comparison between digitized and predicted boulders in (b–d).

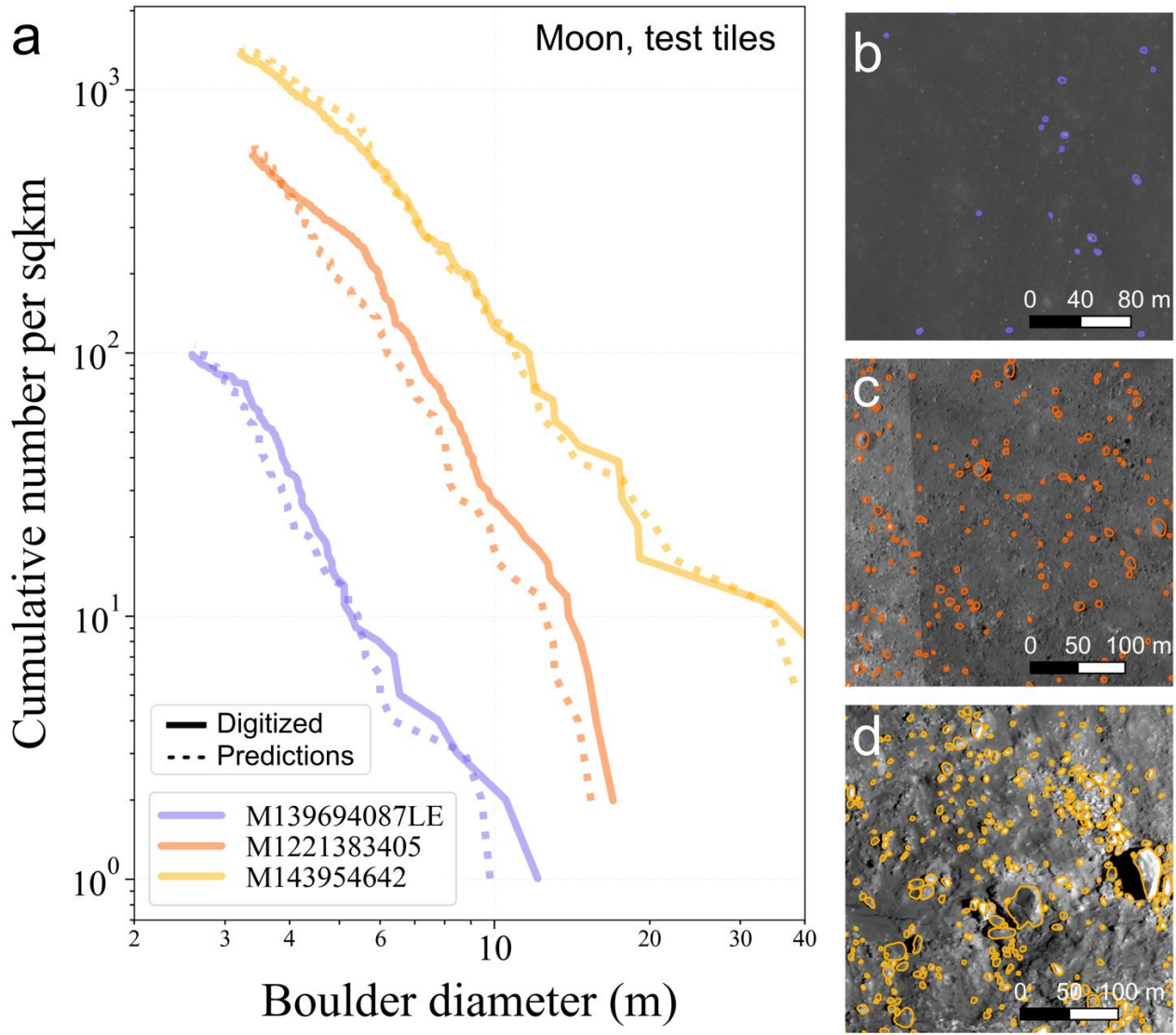

**Figure 5.** (a) Boulder-size frequency distribution (BSFD) curves for manually digitized (solid lines) and predicted (dashed lines) boulders at the lunar locations shown in (b-d). Only boulders with surface areas > 6×6 pixels are shown (similar to manual mapping methodology, but with one added pixel in each direction to minimize potential threshold effects); variability on the smallest mapped boulders results from variable pixel resolution among sites. Example close-up predicted boulder outlines at a randomly picked test tile within (b) M139694087LE (c) M1221383405, and (d) M143954642. Supplementary file Animation S2 shows a comparison between digitized and predicted boulders in (b–d).

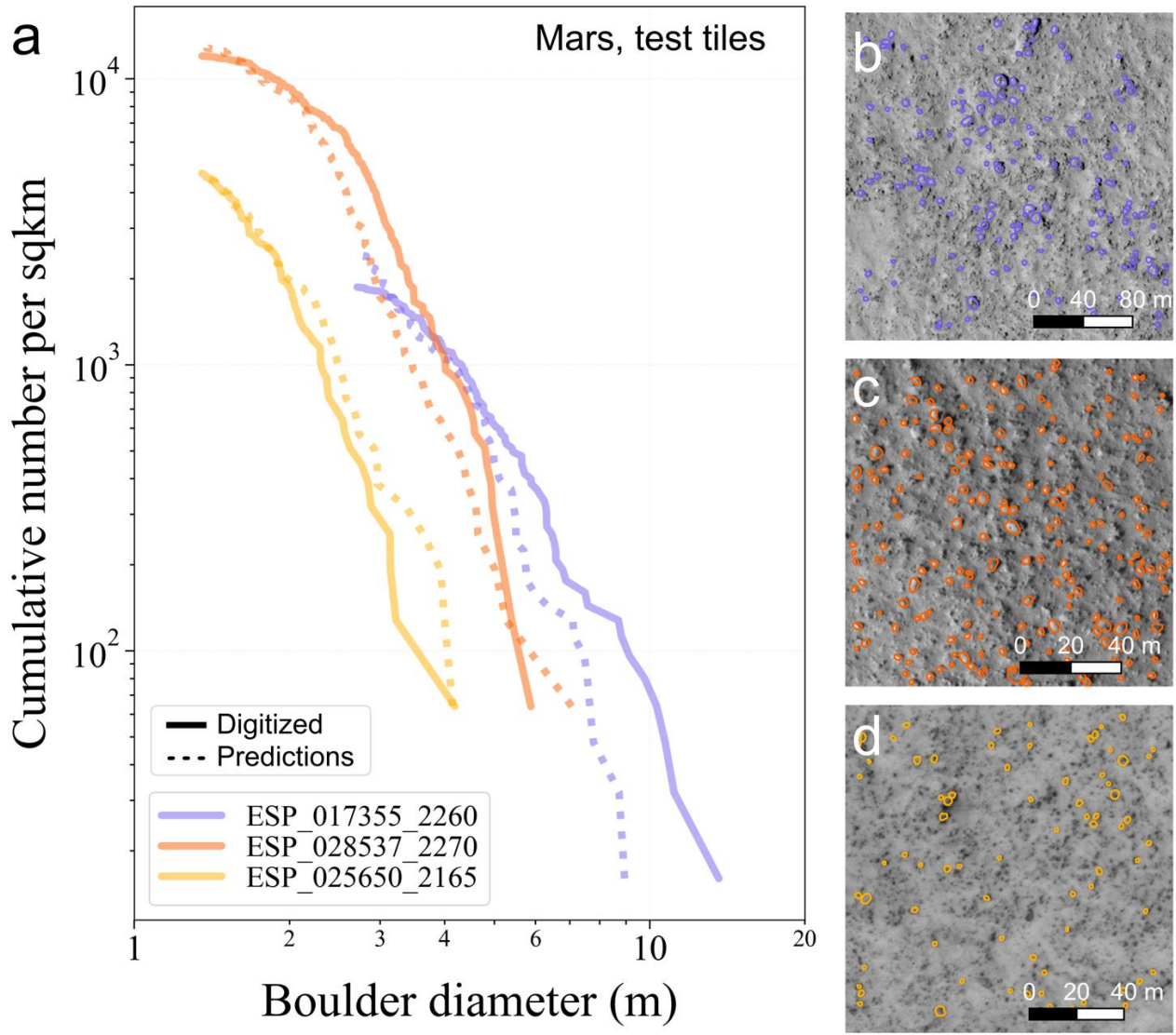

**Figure 6.** (a) Boulder-size frequency distribution (BSFD) curves for manually digitized (solid lines) and predicted (dashed lines) boulders at the martian locations shown in (b-d). Only boulders with surface areas > 6×6 pixels are shown (similar to manual mapping methodology, but with one added pixel in each direction to minimize potential threshold effects); variability on the smallest mapped boulders results from variable pixel resolution among sites. Example close-up predicted boulder outlines at a randomly picked test tile within (b) ESP_017355_2260 (c) ESP_028537_2270, and (d) ESP_025650_2165. Supplementary file Animation S3 shows a comparison between digitized and predicted boulders in (b–d).

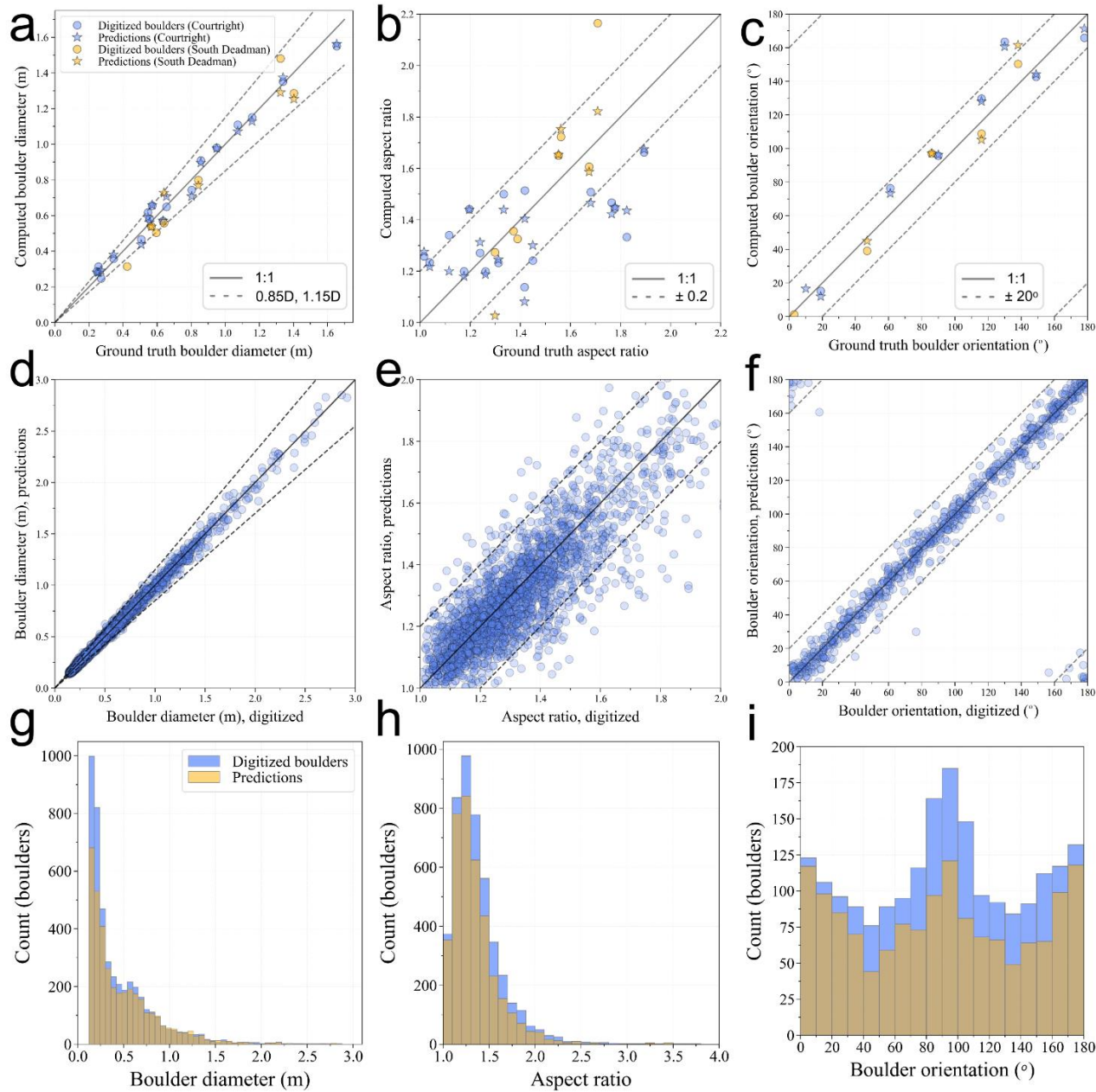

**Figure 7.** Scatter-plot comparison between ground truth, manual mapping, and model predictions of (a) equivalent boulder diameter, (b) boulder aspect ratio, and (c) boulder orientation. Scatter-plot comparison between manual mapping and model predictions for (d) equivalent boulder diameter, (e) boulder aspect ratio, (f) and boulder orientation at Courtright Reservoir. Only pairs IoU ≥ 0.75 are matched together. Histograms of manually mapped and predicted (g) equivalent boulder diameters (bin size = 0.06 m), (h) boulder aspect ratios (bin size = 0.1), and (i) boulder orientations (bin size = 10°) at Courtright Reservoir. Boulder orientations (a,f,i) are only computed for boulders having aspect ratio ≥ 1.35.

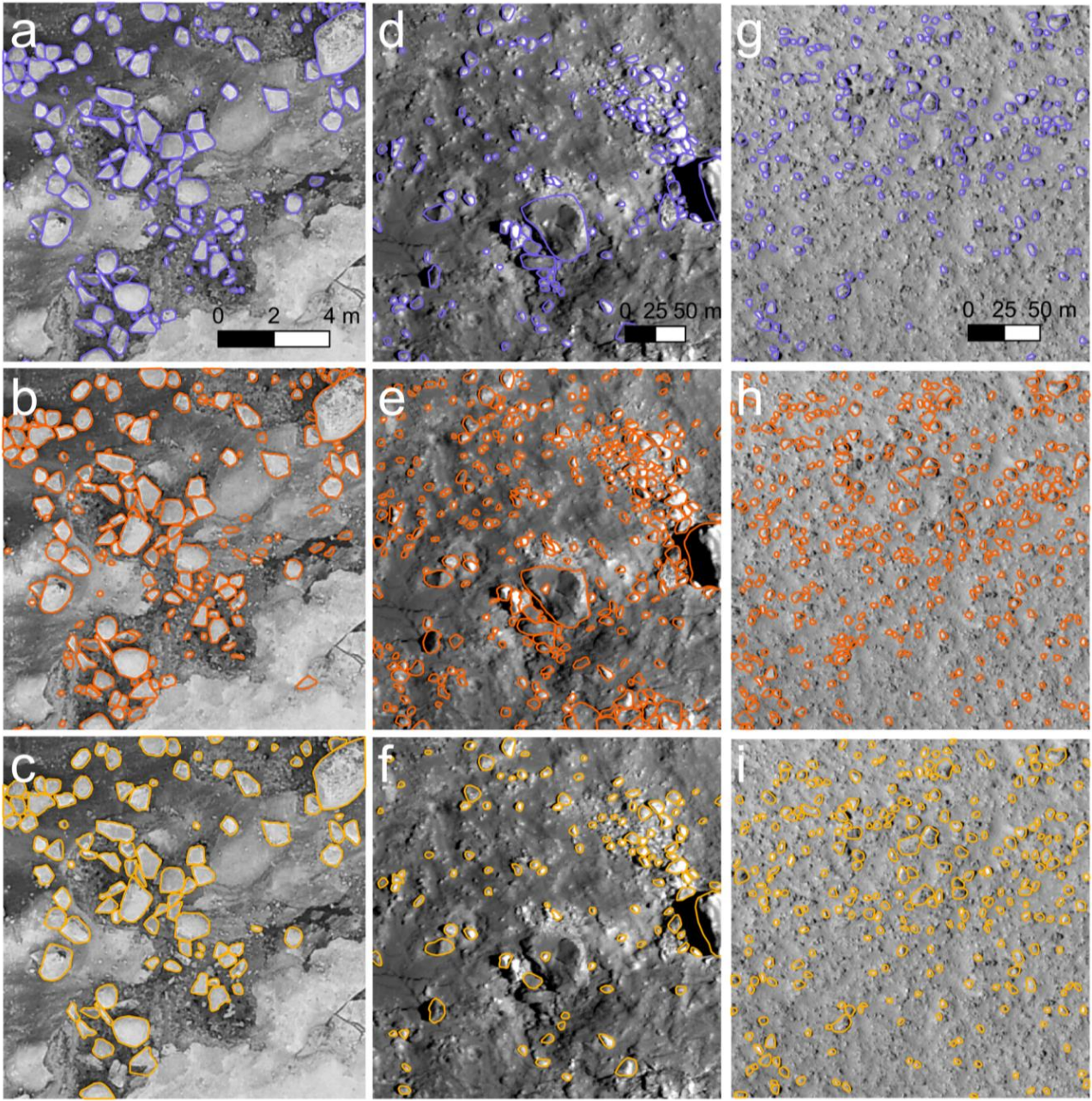

**Figure 8.** Variance in manually mapped boulder outlines among three human mappers for three challenging mapping tiles (512x512 pixels) (a–c) on Earth (Courtright Reservoir), (d-f) the Moon, (M143954642), and (g–i) on Mars (ESP_017355_2260). Each color indicates mapping from a single human annotator. Only boulders with surface areas > 6×6 pixels are shown.

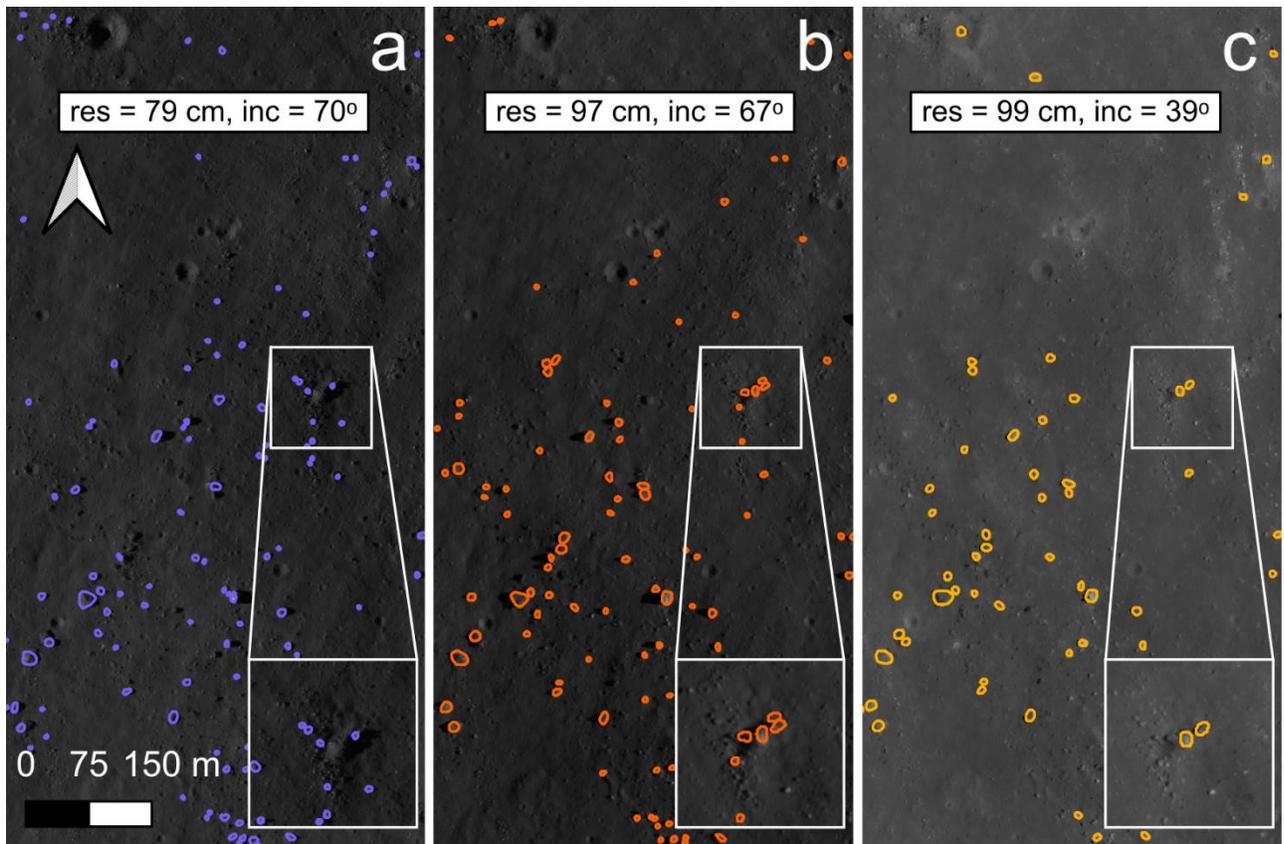

**Figure 9.** Boulder predictions derived from three NAC images (M1218745643LE, M1123377165RE, M1151645127RE, left to right) of the northern rim of Censorinus impact crater (0.338°S, 32.661°E). The resolution (*res*) and incidence angles (*inc*, reported relative to vertical, i.e., nadir corresponds to 0°) of each image are specified in each panel. The white rectangles in the middle right of images are zoomed in in the lower right corners for better visibility. Only boulders with surface areas > 6×6 pixels are shown.

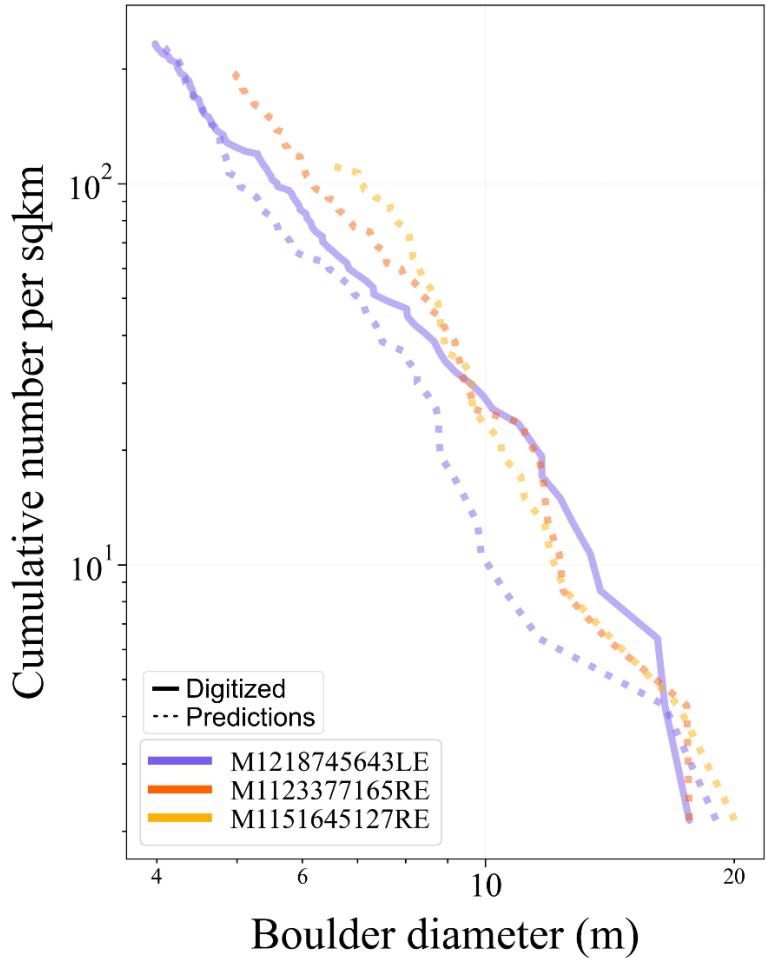

**Figure 10.** Boulder-size frequency distributions (BSFD) of manually digitized (solid line, only computed for the highest resolution NAC image, M1218745643LE) and predicted (dashed lines) boulders at the location shown in Figure 9 (northern rim of Censorinus impact crater, 0.338°S, 32.661°E). The same colors are used in Figures 9 and 10. Only boulders with surface areas > 6×6 pixels are shown; variability on the smallest mapped boulders results from variable pixel resolution among images.

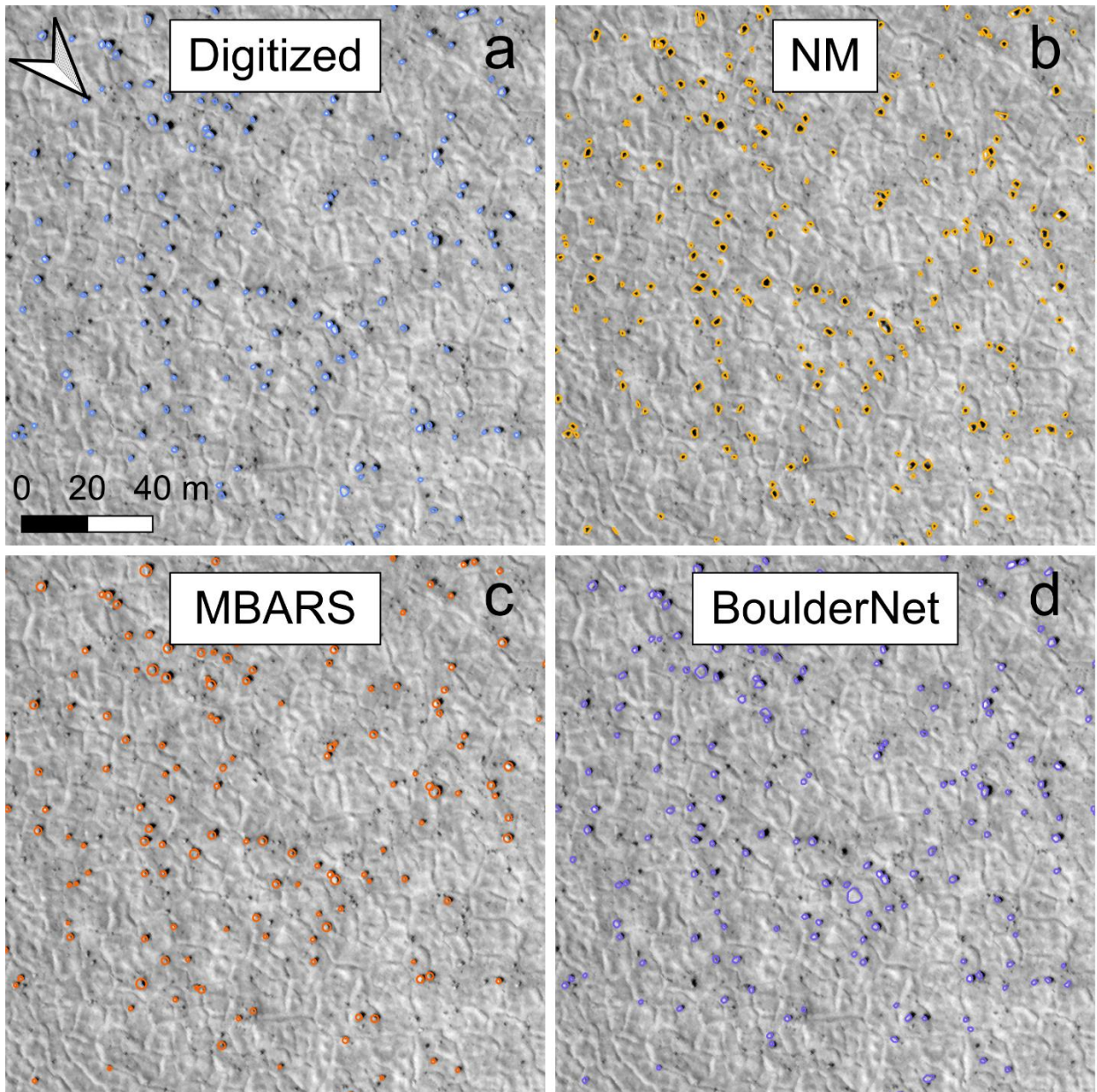

**Figure 11.** Boulder outlines of (a) manually digitized and (b–d) predictions from three different existing boulder detection algorithms – (b) the NM-method, (c) MBARS and (d) our detection algorithm, BoulderNet. Only boulders with areas more than 10 pixels are shown.

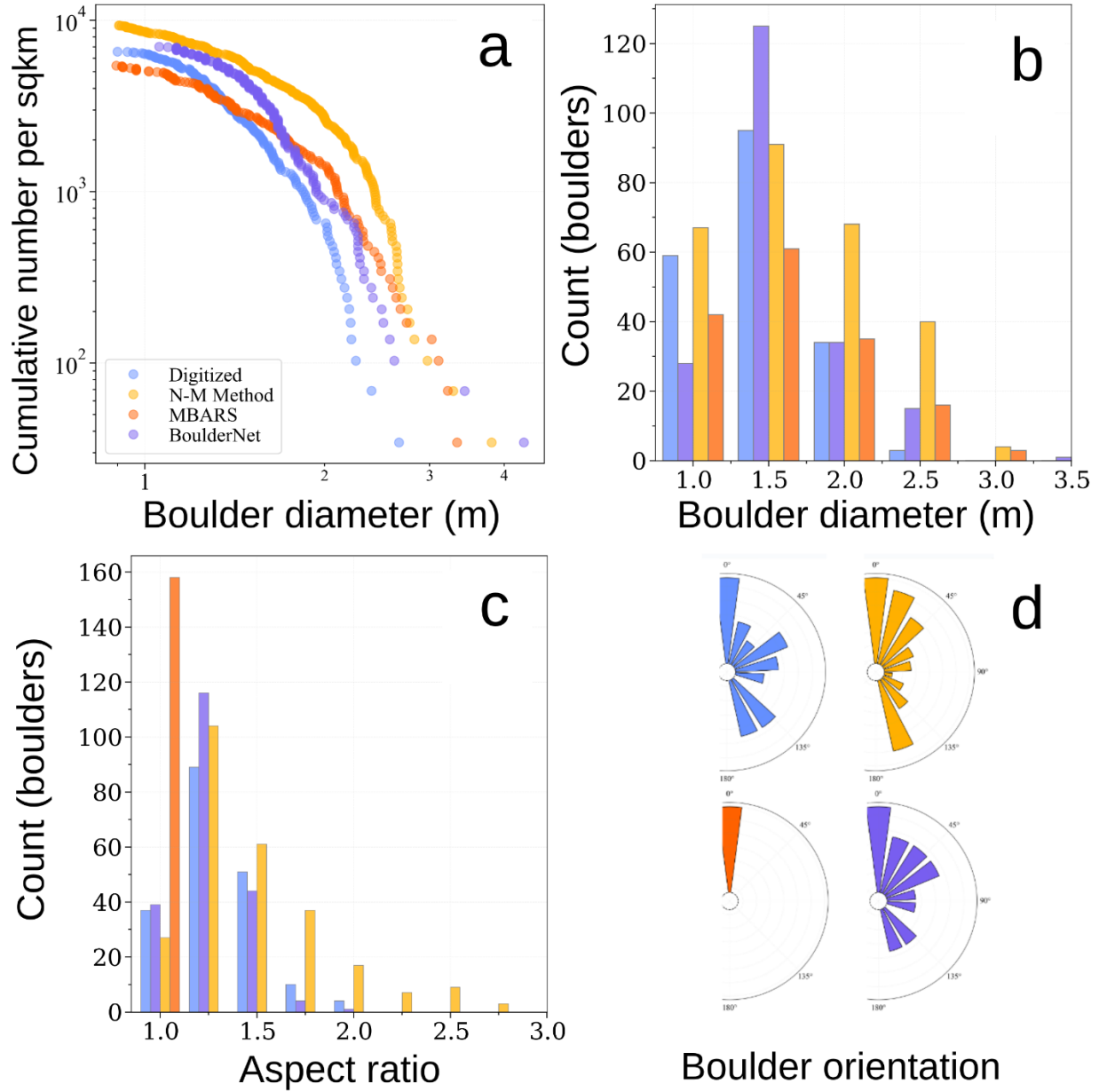

**Figure 12.** Comparison of (a) boulder-size frequency distributions (BSFD), histograms of (b) equivalent boulder diameter (bin size = 0.50 m) and (c) aspect ratio (bin size = 0.25), and (d) angular distribution of boulder orientation (bin size = 20°) for manually digitized and predictions made from the three boulder detection algorithms of Figure 11. Only predictions intersecting manually digitized boulders are used in (d).

# Tables

**Table 1.** Boulder database (as of 25/04/2023). LRO, NAC, MRO, and HiRISE stand respectively for Lunar Reconnaissance Orbiter, Narrow Angle Camera, Mars Reconnaissance Orbiter, and High-Resolution Imaging Science Experiment. Incidence angles are reported relative to vertical (i.e., nadir corresponds to 0º).

| Planetary body | Instrument | Image ID | n tiles mapped | n boulders mapped | Resolution (m/pix) | Incidence angle (°) |
|---|---|---|---|---|---|---|
| Earth | Drone | Courtright Reservoir | 180 | 4750 | 0.03 | 33 |
| Earth | Drone | South Deadman | 3 | 1552 | 0.03 | 40 |
| Earth | Drone | Stonegarden | 65 | 5943 | 0.06 | 32 |
| Mars | MRO HiRISE | ESP_017355_2260 | 20 | 4509 | 0.50 | 42 |
| Mars | MRO HiRISE | ESP_025650_2165 | 10 | 665 | 0.25 | 38 |
| Mars | MRO HiRISE | ESP_028537_2270 | 10 | 1619 | 0.25 | 58 |
| Mars | MRO HiRISE | ESP_037018_2170 | 10 | 189 | 0.60 | 53 |
| Moon | LRO NAC | M104827900 | 16 | 726 | 0.74 | 53 |
| Moon | LRO NAC | M1097858669 | 5 | 68 | 1.19 | 55 |
| Moon | LRO NAC | M1098366480 | 15 | 318 | 0.95 | 49 |
| Moon | LRO NAC | M1113133069 | 16 | 571 | 0.73 | 61 |
| Moon | LRO NAC | M1116885123 | 10 | 374 | 1.56 | 52 |
| Moon | LRO NAC | M1221383405 | 97 | 6832 | 0.63 | 41 |
| Moon | LRO NAC | M123784480 | 16 | 475 | 0.57 | 32 |
| Moon | LRO NAC | M1288478295 | 8 | 146 | 0.74 | 51 |
| Moon | LRO NAC | M139694087 | 338 | 6026 | 0.48 | 9 |
| Moon | LRO NAC | M141573316 | 13 | 402 | 0.58 | 25 |
| Moon | LRO NAC | M143954642 | 23 | 1093 | 0.60 | 42 |
| Moon | LRO NAC | M169764515 | 18 | 792 | 0.47 | 28 |
| Moon | LRO NAC | M174285569 | 35 | 908 | 0.60 | 31 |
| Moon | LRO NAC | M183960297 | 10 | 546 | 0.72 | 47 |
| Moon | LRO NAC | M186654867 | 18 | 472 | 0.62 | 55 |
| Moon | LRO NAC | M190294287 | 17 | 564 | 1.02 | 41 |
| **Total** | | | **953** | **39,540** | | |

**Table 2.** Aggregated average precision (AP) and recall (AR) in percent for a IoU thresholds of 50 and 75% obtained from the best model across all images in our dataset. Model performances for "raw" predictions can be found in Supplementary Table 2 (Supplementary Information).

| \multicolumn{4}{c}{*Test dataset*} |
|---|---|---|---|

| AP50 | AR50 | AP75 | AR75 |
|---|---|---|---|
| 72 | 64 | 40 | 35 |
| \multicolumn{4}{c}{*Validation dataset*} | | | |
| AP50 | AR50 | AP75 | AR75 |
| 72 | 61 | 40 | 35 |

**Table 3.** Average precision (AP) and recall (AR) values in percent for IoU thresholds of 50 and 75% obtained from the best model for locations shown in Figures 4–6.

| | **Earth** (Figure 4a) | | | | |
|---|---|---|---|---|---|
| | size range | AP50 | AR50 | AP75 | AR75 |
| **Courtright Reservoir** | all | 87 | 85 | 72 | 70 |
| | small | 74 | 76 | 52 | 54 |
| | medium | 96 | 93 | 88 | 85 |
| | large | 100 | 85 | 100 | 85 |
| **South Deadman** | all | 77 | 61 | 43 | 34 |
| | small | 70 | 53 | 34 | 26 |
| | medium | 74 | 67 | 51 | 47 |
| | large | 80 | 54 | 48 | 32 |
| **Stonegarden** | all | 74 | 65 | 48 | 42 |
| | small | 66 | 59 | 41 | 37 |
| | medium | 83 | 69 | 57 | 47 |
| | large | 85 | 74 | 80 | 70 |
| | **Moon** (Figure 5a) | | | | |
| **M139694087LE** | all | 66 | 53 | 28 | 22 |
| | small | 64 | 50 | 28 | 22 |
| | medium | 33 | 50 | - | - |
| | large | - | - | - | - |
| **M1221383405** | all | 70 | 61 | 35 | 30 |
| | small | 68 | 61 | 31 | 28 |
| | medium | 100 | 60 | 100 | 60 |
| | large | - | - | - | - |
| **M143954642** | all | 66 | 73 | 31 | 35 |
| | small | 64 | 72 | 31 | 34 |
| | medium | 65 | 65 | 35 | 35 |
| | large | 67 | 100 | 33 | 50 |
| | **Mars** (Figure 6a) | | | | |
| **ESP_017355_2260** | all | 60 | 58 | 21 | 20 |
| | small | 59 | 61 | 20 | 20 |
| | medium | - | - | - | - |
| | large | - | - | - | - |
| **ESP_028537_2270** | all | 71 | 68 | 23 | 22 |
| | small | 70 | 69 | 22 | 21 |
| | medium | 80 | 40 | 40 | 20 |

|  | large | - | - | - | - |
|---|---|---|---|---|---|
| **ESP_025650_2165** | all | 38 | 18 | 12 | 14 |
|  | small | 38 | 18 | 12 | 14 |
|  | medium | - | - | - | - |
|  | large | - | - | - | - |

**Table 4.** Quantitative comparison of manual boulder mapping among three human annotators at three locations (tiles in Figure 8). The table depicts the average precisions and recalls for IoU $\geq$ 0.5 (i.e., AP50 and AR50) and IoU $\geq$ 0.75 (in parenthesis). Only boulders with areas larger than 36 pixels are included in the calculation. See Table 1 for more image resolution and illumination conditions.

|  | Courtright Reservoir | | M143954642 | | ESP_017355_2260 | |
|---|---|---|---|---|---|---|
|  | AP50 (AP75) | AR50 (AR75) | AP50 (AP75) | AR50 (AR75) | AP50 (AP75) | AR50 (AR75) |
| Mapper 1 vs Mapper 2 | 72 (63) | 89 (77) | 34 (8) | 76 (17) | 45 (15) | 87 (30) |
| Mapper 1 vs Mapper 3 | 99 (95) | 65 (62) | 83 (48) | 61 (35) | 63 (19) | 88 (27) |
| Mapper 2 vs Mapper 3 | 99 (97) | 53 (51) | 92 (39) | 30 (13) | 81 (46) | 58 (33) |
| Average | 90 (85) | 69 (63) | 70 (32) | 56 (22) | 63 (27) | 78 (30) |
| Standard deviation | 13 (16) | 15 (11) | 25 (17) | 19 (10) | 15 (14) | 14 (2) |